\documentclass{jps-cp}

\usepackage{txfonts} 

\title{Superconductivity in Nb$_{5}$Ir$_{3-x}$Pt$_{x}$O}

\author{Jiro \textsc{Kitagawa} and Shusuke \textsc{Hamamoto}}

\inst{Department of Electrical Engineering, Faculty of Engineering, Fukuoka Institute of Technology, 3-30-1 Wajiro-higashi, Higashi-ku, Fukuoka 811-0295, Japan}

\email{j-kitagawa@fit.ac.jp}

\recdate{September 3, 2019}

\abst{We have investigated the superconducting critical temperature $T_\mathrm{c}$ of solid solution Nb$_{5}$Ir$_{3-x}$Pt$_{x}$O with the Ti$_{5}$Ga$_{4}$-type structure. The both end-members of Nb$_{5}$Ir$_{3}$O and Nb$_{5}$Pt$_{3}$O are reported to be superconductors with $T_\mathrm{c}$ of 10.5 K and 3.8 K, respectively. Particularly Nb$_{5}$Ir$_{3}$O is considered as a two-gap superconductor. The entire series of alloy hold the Ti$_{5}$Ga$_{4}$-type structure and show the linear $x$ dependence of lattice parameters. On the other hand, a nonlinear $T_\mathrm{c}$ vs $x$ plot was obtained, suggesting that $T_\mathrm{c}$ is not determined by only lattice parameters. The experimental result is discussed based on the Matthias rule and compared with those of the other two-gap superconductors such as MgB$_{2}$, Mo$_{8}$Ga$_{41}$ and Nb$_{3}$Sn.}

\kword{superconductivity, Ti$_{5}$Ga$_{4}$-type, Nb$_{5}$Ir$_{3}$O}

\begin{document}
\maketitle

\section{Introduction}
In the Mn$_{5}$Si$_{3}$-type structure with the space group P6$_{3}$/mcm (No.193), the Mn sites of 4d and 6g Wyckoff symmetries are occupied by early transition metals, rare earth or alkaline earth elements, and the Si sites of 6g Wyckoff symmetry by metalloid elements or post-transition metals.
The Mn$_{5}$Si$_{3}$-type structure possesses the interstitial 2b site, allowing the addition of light elements such as oxygen, boron and carbon with no change of the space group.
Such an ordered variant of Mn$_{5}$Si$_{3}$-type structure is called as the Ti$_{5}$Ga$_{4}$ or Hf$_{5}$CuSn$_{3}$-type structure.
Figures 1(a) and 1(b) show the crystal structure of Ti$_{5}$Ga$_{4}$-type Nb$_{5}$Ir$_{3}$O investigated in this study.
The oxygen atom is surrounded by Nb2 atoms in the octahedral site (6g site), which forms a face-sharing Nb2$_{6}$ chain along the $c$-axis.
Another octahedral Ir atoms enclose the Nb1 atom (4d site) also forming a one-dimensional atomic chain along the $c$-axis.
Therefore an expansion of lattice parameter $a$ would enhance the one-dimensional nature of octahedral Nb2$_{6}$ and Nb1 atomic chains.

\begin{figure}[tbh]
\begin{center}
\includegraphics[width=13cm]{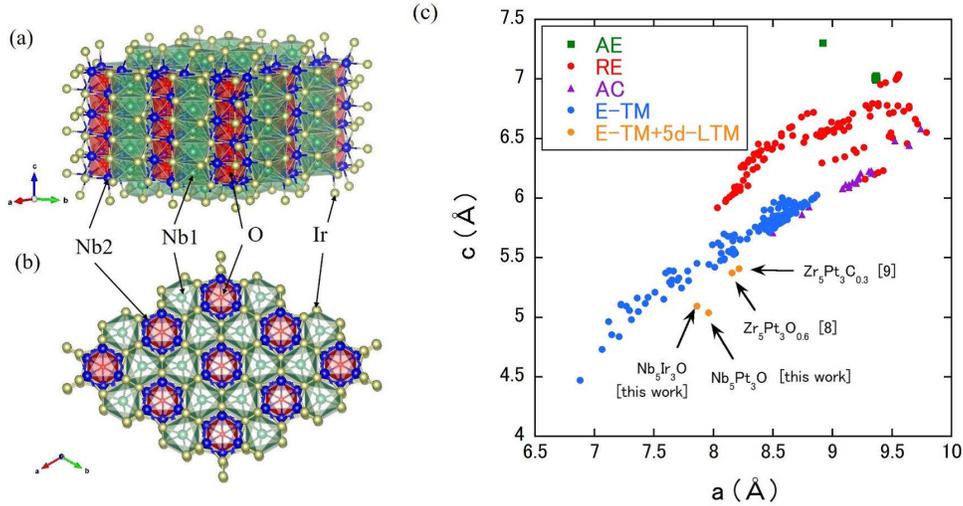}
\caption{(a)(b) Crystal structure of Nb$_{5}$Ir$_{3}$O. (c) $c$ vs $a$ plot of Ti$_{5}$Ga$_{4}$-type compounds. Except E-TM+5$d$-LTM group, the data of commercial crystallographic database CRYSTMET are employed.}
\end{center}
\label{f1}
\end{figure}

The anisotropy parameter of $c/a$ emphasizes the crystallographic feature of Ti$_{5}$Ga$_{4}$-type compounds (see Fig.\ 1(c)).
We have classified the compounds into four groups AE, RE, AC and E-TM, where the Mn-sites are occupied by alkaline earth, rare earth, actinide and early transition metal elements, respectively.
On going from AE, RE and E-TM (AC), $c/a$ seems to decrease, suggesting an important role of atomic size of the Mn-site element for determining the anisotropy.
In the case of AE and RE, the rather large values of both lattice parameters are probably due to a weak bonding between the parent Mn$_{5}$Si$_{3}$-type compound and added interstitial elements, which is caused by the rather poor electron counts in parent compound.
We separate the E-TM group compounds in which 5$d$ late transition metals like Ir and Pt occupy the Si sites from the E-TM group, and denote E-TM+5$d$-LTM in Fig.\ 1(c).
The E-TM+5$d$-LTM group possesses $c/a$ smaller than that of E-TM group, which means the enhancement of anisotropy.
It should be noted that all compounds in the E-TM+5$d$-LTM group are superconductors as mentioned below. 

Although physical properties of many Ti$_{5}$Ga$_{4}$-type compounds are studied\cite{Zheng:JALCOM2002,Surgers:PRB2003,Mar:CM2006,Goruganti:JAP2009}, the superconductivity is reported only in several compounds\cite{Zhang:npjQM2017,Cort:JLTP1982,Bortolozo:JAP2012,Hamamoto:MRX2018,Renosto:arxiv2018} such as Nb$_{5}$Ir$_{3}$O, Nb$_{5}$Pt$_{3}$O, Nb$_{5}$Ge$_{3}$C$_{0.3}$, Zr$_{5}$Pt$_{3}$O$_{x}$ and Zr$_{5}$Pt$_{3}$C$_{x}$.
For Nb$_{5}$Ir$_{3}$O, the non oxygen-doped Nb$_{5}$Ir$_{3}$ is a superconductor\cite{Zhang:npjQM2017} with the critical temperature $T_\mathrm{c}=$ 9.3 K, and by increasing the oxygen concentration, $T_\mathrm{c}$ progressively increases to 10.5 K.
In this case, the oxygen addition leads to the shrinkage of $c$ and the expansion of $a$, which is regarded as the enhancement of one-dimensionality of Nb2$_{6}$ and Nb1 atomic chains. 
The temperature dependence of specific heat of Nb$_{5}$Ir$_{3}$O cannot be reproduced by a single exponential temperature dependence, but explained by a two-gap model\cite{Zhang:npjQM2017}.
The bulk superconductivity with $T_\mathrm{c}$ of 3.8 K has been confirmed for Nb$_{5}$Pt$_{3}$O by measuring the specific heat, which seems to follow a single exponential temperature dependence just below $T_\mathrm{c}$ but shows the deviation at low temperature\cite{Cort:JLTP1982}.
In this study, Pt-substitution effect on $T_\mathrm{c}$ is investigated in the solid solution system Nb$_{5}$Ir$_{3-x}$Pt$_{x}$O to further elucidate the superconductivity of Ti$_{5}$Ga$_{4}$-type compounds.

\section{Materials and Methods}
Polycrystalline samples were prepared using Nb powder (99.9\%), Nb$_{2}$O$_{5}$ powder (99.9\%), Ir powder (99.99\%) and Pt wire (99.9\%).
Nb$_{5}$Ir$_{3-x}$O was initially prepared by arc melting a pellet made by pressing a homogenized mixture of Nb, Nb$_{2}$O$_{5}$ and Ir powders.
Then Nb$_{5}$Ir$_{3-x}$O was remelted with added Pt wire to form the stoichiometric composition.
The samples were remelted several times to ensure the homogeneity of the samples.
The weight loss during the arc melting was negligible.
Each as-cast sample was annealed in an evacuated quartz tube at 800 $^{\circ}$C for 4 days.
A powder X-ray diffractometer (Shimadzu, XRD-7000L) with Cu-K$\alpha$ radiation was used to measure the X-ray diffraction (XRD) patterns of prepared samples. 
The temperature dependences of ac magnetic susceptibility $\chi_{ac}$ (T) and electrical resistivity $\rho$ (T), between 2.8 K and 300 K, were measured using a closed-cycle He gas cryostat.
The mutual inductance method with an alternating filed of 5 Oe and 800 Hz was employed to measure $\chi_{ac}$ (T).

\section{Results and Discussion}
Figure 2(a) shows the XRD patterns of representative samples, which can be indexed by the Ti$_{5}$Ga$_{4}$-type structure except the minor impurity peaks indicated by the asterisks.
The impurity phase is assigned as Nb$_{3}$Ir$_{2}$-Nb$_{3}$Pt$_{2}$ solid solution.
The amount of Nb$_{3}$Ir$_{2}$, which is a superconductor\cite{Koch:PRB1971} with $T_\mathrm{c}\sim$ 2.3 K, in $x=$ 0 sample is rather large, but the $T_\mathrm{c}$ is below the our lowest measurement temperature, and the impurity phase would have no influence on our experimental results. 
With increasing $x$, the peak positions with Miller indices ($h$ $k$ 0) and (0 0 $l$) shift to lower and higher 2$\theta$ angles, respectively.
This implies that the lattice parameter $a$ ($c$) increases (decreases) by substituting Pt into Ir.
The lattice parameters are obtained by the least square method\cite{Tsubota:SR2017} and the $x$ dependence of these parameters are shown in Fig.\ 2(b).
By increasing $x$, $a$ ($c$) linearly expands (shrinks), which means an enhanced one-dimensional nature of octahedral Nb2$_{6}$ and Nb1 atomic chains.

\begin{figure}[tbh]
\begin{center}
\includegraphics[width=14cm]{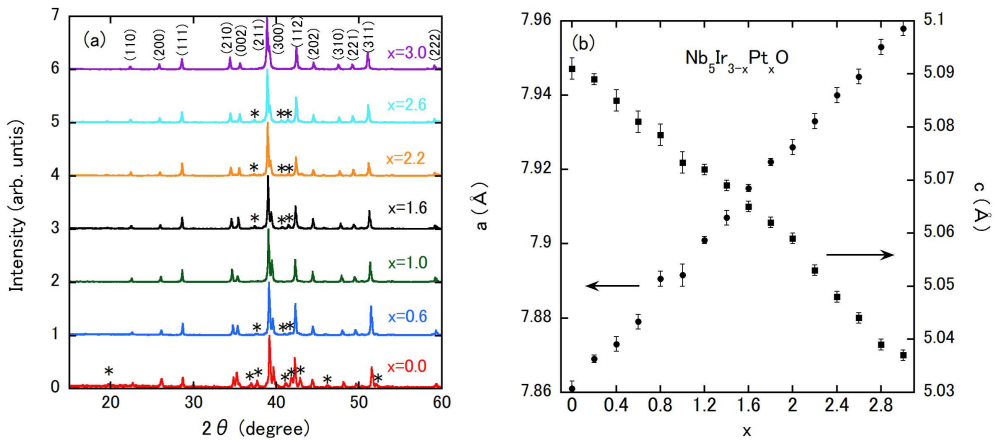}
\caption{(a) XRD patterns of Nb$_{5}$Ir$_{3-x}$Pt$_{x}$O. The origin of each pattern is shifted by an integer value for clarity. (b) $x$ dependence of lattice parameters of Nb$_{5}$Ir$_{3-x}$Pt$_{x}$O.}
\end{center}
\label{f2}
\end{figure}

$\chi_{ac}$ (T) of Nb$_{5}$Ir$_{3-x}$Pt$_{x}$O with 0.0$\leq x \leq$ 1.6 and those with 1.8$\leq x \leq$ 3.0 are shown in Figs.\ 3(a) and 3(b), respectively.
Each sample exhibits diamagnetic signal and the $T_\mathrm{c}$ was determined as being the intercept of the linearly extrapolated diamagnetic slope with the normal state signal (see the broken lines in the Fig.\ 3(a)).
$T_\mathrm{c}$'s of Nb$_{5}$Ir$_{3}$O and Nb$_{5}$Pt$_{3}$O are approximately 10.1 K and 4.3 K, respectively, which are in agreement with the literature values\cite{Zhang:npjQM2017,Cort:JLTP1982}.
As $x$ is increased, $T_\mathrm{c}$ does not shift for the sample with $x \leq$ 0.6 and starts to decrease at $x \geq$ 0.6.

\begin{figure}
\begin{center}
\includegraphics[width=15cm]{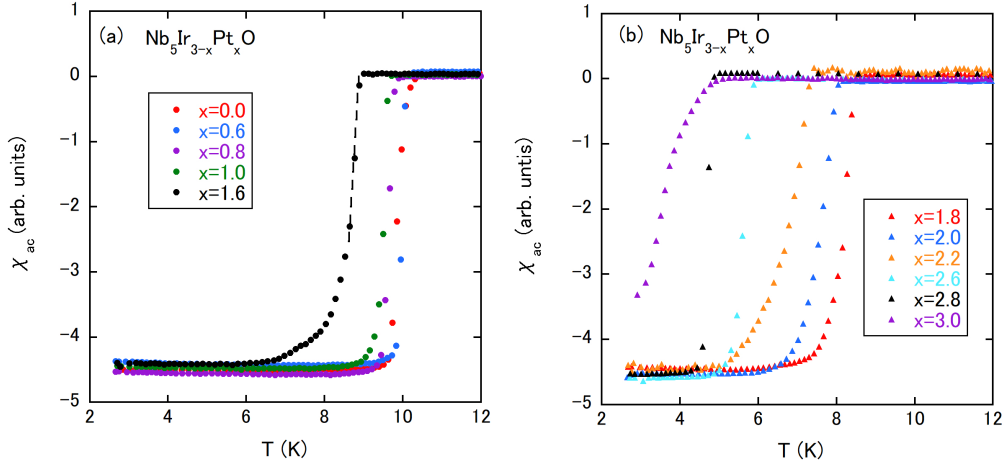}
\caption{(a) Temperature dependence of $\chi_{ac}$ for Nb$_{5}$Ir$_{3-x}$Pt$_{x}$O with $x$=0.0, 0.6, 0.8, 1.0 and 1.6. (b) Temperature dependence of $\chi_{ac}$ for Nb$_{5}$Ir$_{3-x}$Pt$_{x}$O with $x$=1.8, 2.0, 2.2, 2.6, 2.8 and 3.0.}
\end{center}
\label{f3}
\end{figure}

Figure 4 shows $\rho$ (T) of representative Nb$_{5}$Ir$_{3-x}$Pt$_{x}$O samples, which are normalized by the room-temperature values listed in Table I.
Except Nb$_{5}$Pt$_{3}$O, each sample shows the zero resistivity below $T_\mathrm{c}$.
For Nb$_{5}$Pt$_{3}$O with low $T_\mathrm{c}$, zero resistivity could not be observed at the lowest achievable temperature.
All $\rho$ (T) curves deviate from the linearity above $T_\mathrm{c}$ (see Fig.\ 4(b)), which is also observed\cite{Woodard:PR1964,Hiroi:PRB2007} in A15 superconductors like Nb$_{3}$Sn or a pyrochloa superconductor of KOs$_{2}$O$_{6}$.
An additional scattering source\cite{Woodard:PR1964,Hiroi:PRB2007} is responsible for the deviation, and Woodward and Cody\cite{Woodard:PR1964} have presented a well-known empirical formula as follows:
\begin{equation}
 \rho=\rho_{0}+AT+\rho_{1}exp(-\frac{T_{0}}{T})
\label{equ:WC}
\end{equation}
, where the first term means a residual resistivity, the second one phonon part of $\rho$ and the third one describes anomalous temperature dependence.
We have also fitted $\rho$ (T) in Fig.\ 4(b) using eq. (1), and the well reproducibilities as depicted by the solid curves are obtained.
As can be seen from Table I summarizing the fitting parameters, $\rho_{1}$ roughly corresponds to $\rho$(RT)$-\rho_{0}$.
For the almost samples, the third term in eq.(1) dominates over the second one, and $T_{0}$ does not largely change by the Pt substitution. 
We note that the deviation from the linearity is commonly observed in the Ti$_{5}$Ga$_{4}$-type superconductors\cite{Hamamoto:MRX2018,Bortolozo:PhysicaB} such as Zr$_{5}$Pt$_{3}$O$_{x}$ and Nb$_{5}$Ge$_{3}$C$_{0.3}$.  

\begin{figure}
\begin{center}
\includegraphics[width=14cm]{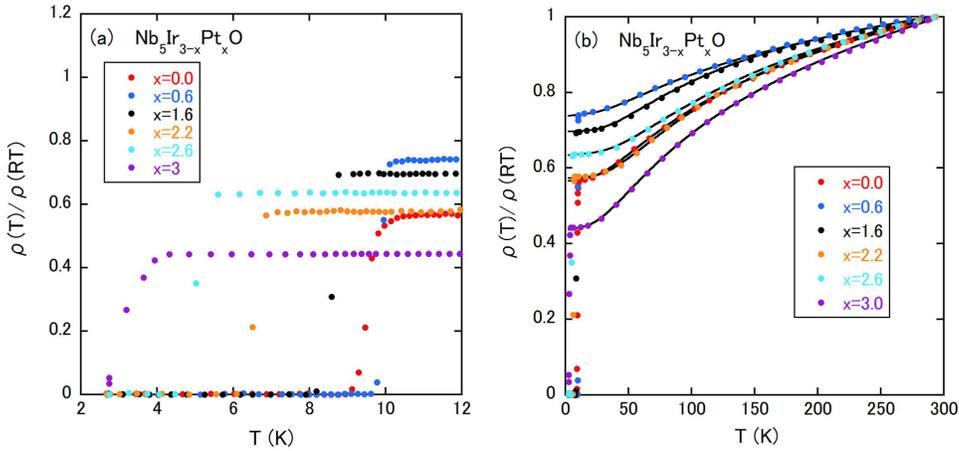}
\caption{Temperature dependence of $\rho$ for Nb$_{5}$Ir$_{3-x}$Pt$_{x}$O with $x$=0.0, 0.6, 1.6, 2.2, 2.6 and 3.0 at (a) low temperature and (b) between 2.8 and 300 K. The solid curves in (b) are $\rho$ (T) calculated by eq.(1).}
\end{center}
\label{f4}
\end{figure}

\begin{table}[tbh]
\begin{center}
\caption{Room-temperature $\rho$ and parameters obtained by the fitting using eq.(1) for Nb$_{5}$Ir$_{3-x}$Pt$_{x}$O with $x$=0.0, 0.6, 1.6, 2.2, 2.6 and 3.0.}
\label{t1}
\begin{tabular}{llllll}
\hline
$x$ & $\rho$ (RT) ($\mu\Omega$ cm) & $\rho_{0}$ ($\mu\Omega$ cm) & $A$ ($\mu\Omega$ cm/K) & $\rho_{1}$ ($\mu\Omega$ cm) & $T_{0}$ (K)  \\
\hline
0.0 & 271 & 153 & 0.13 & 115 & 109 \\
0.6 & 186 & 137 & 0.067 & 44 & 117 \\
1.6 & 162 & 113 & 0.024 & 64 & 121 \\
2.2 & 337 & 193 & 0.074 & 189 & 129 \\
2.6 & 150 & 95 & 0.046 & 68 & 144 \\
3.0 & 503 & 219 & 0.32 & 265 & 101 \\
\hline
\end{tabular}
\end{center}
\end{table}

Figure 5(a) shows the $x$ dependence of $T_\mathrm{c}$, which is contrasted with the linear $x$ dependence of lattice parameters.
Although the latter behavior may support the enhancement of one dimensional nature of Nb2$_{6}$ and Nb1 atomic chains with increasing $x$, leading to an increase of $T_\mathrm{c}$ as observed in Nb$_{5}$Ir$_{3}$O$_{x}$ (see the Introduction), Nb$_{5}$Ir$_{3-x}$Pt$_{x}$O exhibits the opposite behavior.  
The contrasted results between Nb$_{5}$Ir$_{3-x}$Pt$_{x}$O and Nb$_{5}$Ir$_{3}$O$_{x}$ indicates that the superconductivity of Nb$_{5}$Ir$_{3}$O is robust against the substitution of Pt into Ir and is not largely affected by only the crystal structure parameters. 
The electronic specific heat coefficient $\gamma$ of the normal state reflects the electronic density of states at the Fermi level, which is responsible for the magnitude of $T_\mathrm{c}$.
The $\gamma$-values of Nb$_{5}$Ir$_{3}$O and Nb$_{5}$Pt$_{3}$O are extracted to be 41 and 30 mJ/molK$^{2}$, respectively\cite{Cort:JLTP1982}.
Therefore, the electronic density of states at the Fermi level would partially determine $T_\mathrm{c}$.
The valence electron concentration (VEC) per atom is frequently useful tool for investigating the effect of electronic density of states at the Fermi level.
The $T_\mathrm{c}$ vs VEC plot well describes the tendency of $T_\mathrm{c}$ for body-centered-cubic and A15 superconductors, and generally shows a broad maximum at the specific VEC value (Matthias rule)\cite{Matthias:PR1955}.
Employing $T_\mathrm{c}$ data of the Mn$_{5}$Si$_{3}$ or Ti$_{5}$Ga$_{4}$-type superconductors, we made the $T_\mathrm{c}$ vs VEC plot as shown in Fig.\ 5(b).
Assigned VEC of Nb, Zr, Ir, Pt, O and C atoms are 5, 4, 9, 10, 6 and 4, respectively.
The broad maximum seems to exist at Nb$_{5}$Ir$_{3}$O, and the $x$ dependence of $T_\mathrm{c}$ in this study may follow the Matthias rule.
Although the reliability of Matthias rule for two-gap superconductors is not established and further study is needed, we note that Nb$_{3}$Sn which is well known familiar A15 superconductor but recently regarded as a two-gap superconductor\cite{Guritanu:PRB2004} follows the Matthias rule.

\begin{figure}
\begin{center}
\includegraphics[width=15cm]{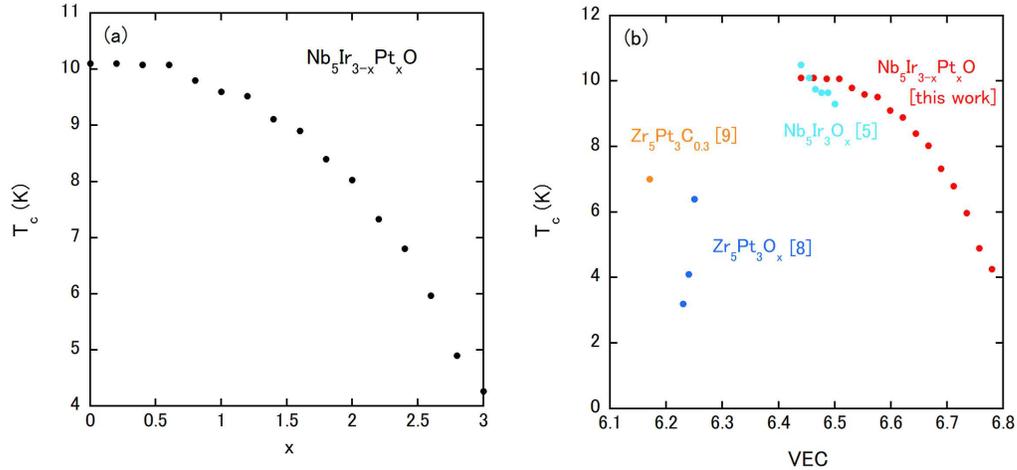}
\end{center}
\caption{(a) $x$ dependence of $T_\mathrm{c}$ determined by $\chi_{ac}$ measurements for Nb$_{5}$Ir$_{3-x}$Pt$_{x}$O. (b) $T_\mathrm{c}$ vs VEC plot for several Mn$_{5}$Si$_{3}$ or Ti$_{5}$Ga$_{4}$-type superconductors.}
\label{f5}
\end{figure}

Focusing on two-gap superconductors, the effects of atomic substitution on $T_\mathrm{c}$ have been reported for several compounds.
Mg$_{1-x}$Al$_{x}$B$_{2}$ or MgB$_{2-x}$C$_{x}$ shows a linear suppression of $T_\mathrm{c}$ by increasing the amount of atomic substitution\cite{Takenobu:PRB2001}.
Especially C-substitution leads to a more rapid decrease of $T_\mathrm{c}$, which means the superconducting properties of MgB$_{2}$ are mainly responsible for boron.
The similar result is confirmed in V-substituted Mo$_{8-x}$V$_{x}$Ga$_{41}$, where the end members of Mo$_{8}$Ga$_{41}$ and V$_{8}$Ga$_{41}$ are 9.7 K two-gap superconductor and non superconductor, respectively\cite{Verchenko:PRB2016}.
On the other hand, Nb$_{3}$Sn provides the non monotonous substitution effect of $T_\mathrm{c}$, especially when the counterpart of end member is also a superconductor.
For example, Nb$_{3}$Sn$_{1-x}$In$_{x}$, in which Nb$_{3}$In is 5 K superconductor, shows the $x$ independent $T_\mathrm{c}$ up to $x \sim$0.12\cite{Otto:ZP1968}.
So our study might indicate that two-gap superconductivity ascribed to the Nb atoms is rather robust against the atomic replacement other than Nb atom, if the counter part of end Nb-compound is also a superconductor.

\section{Summary}
We have determined $T_\mathrm{c}$ of the Ti$_{5}$Ga$_{4}$-type solid solution Nb$_{5}$Ir$_{3-x}$Pt$_{x}$O, in which both end-members are superconductors. 
Especially Nb$_{5}$Ir$_{3}$O is known as a two-gap superconductor with $T_\mathrm{c}=$ 10.5 K. 
Although the lattice parameters linearly depend on $x$, $T_\mathrm{c}$ hardly changes up to $x \sim$ 0.6, suggesting that $T_\mathrm{c}$ is not determined by only lattice parameters.
Combining $T_\mathrm{c}$ data of the other Mn$_{5}$Si$_{3}$ or Ti$_{5}$Ga$_{4}$-type superconductors, the $T_\mathrm{c}$ vs VEC plot is constructed and may be explained by the Matthias rule.
The atomic substitution effect on $T_\mathrm{c}$ is compared among two-gap superconductors such as MgB$_{2}$, Mo$_{8}$Ga$_{41}$ and Nb$_{3}$Sn.
Our result is similar to that of Nb$_{3}$Sn$_{1-x}$In$_{x}$.

\end{document}